\documentclass[usenatbib]{mn2e}
\usepackage{graphicx}
\usepackage{ifthen}

\def\ltsima{$\; \buildrel < \over \sim \;$}
\def\simlt{\lower.5ex\hbox{\ltsima}}
\def\gtsima{$\; \buildrel > \over \sim \;$}
\def\simgt{\lower.5ex\hbox{\gtsima}}
\def\kms{{\rm\,km\,s^{-1}}}

\def\kpc{{\rm\,kpc}}

\def\msun{{\rm\,M_\odot}}
\def\lsun{{\rm\,L_\odot}}

\def\pc{{\rm\,pc}}

\def\AA{$\; \buildrel \circ \over {\rm A}$}

\def\s{\ifmmode \widetilde \else \~\fi}
\def\={\overline}

\def\spose#1{\hbox to 0pt{#1\hss}}

\def\lta{\mathrel{\spose{\lower 3pt\hbox{$\mathchar"218$}}
     \raise 2.0pt\hbox{$\mathchar"13C$}}}
\def\gta{\mathrel{\spose{\lower 3pt\hbox{$\mathchar"218$}}
     \raise 2.0pt\hbox{$\mathchar"13E$}}}
\def\Dt{\spose{\raise 1.5ex\hbox{\hskip3pt$\mathchar"201$}}}    % upper case
\def\dt{\spose{\raise 1.0ex\hbox{\hskip2pt$\mathchar"201$}}}    % lower case

\def\dotsfill{\leaders\hbox to 1em{\hss.\hss}\hfill}

\def\Gyr{{\rm\,Gyr}}
\def\FeH{{\rm[Fe/H]}}

\loadboldmathitalic 
\title[Spectroscopy of Draco~II] {Is Draco II one of the faintest dwarf galaxies? First study from Keck/DEIMOS spectroscopy}

\author[N. F. Martin et al.]{Nicolas F. Martin$^{1,2}$, Marla Geha$^3$, Rodrigo A. Ibata$^1$, Michelle L. M. Collins$^3$, 
\newauthor Benjamin P. M. Laevens$^{1,2}$, Eric F. Bell$^4$, Hans-Walter Rix$^2$, Annette M. N. Ferguson$^5$,
\newauthor  Kenneth C. Chambers$^6$, Richard J. Wainscoat$^6$, Christopher Waters$^{6}$\\
$^1$Observatoire astronomique de Strasbourg, Universit\'e de Strasbourg, CNRS, UMR 7550, 11 rue de l'Universit\'e, F-67000 Strasbourg, France\\
$^2$Max-Planck-Institut f\"ur Astronomie, K\"onigstuhl 17, D-69117 Heidelberg, Germany\\
$^3$Astronomy Department, Yale University, New Haven, CT 06520\\
$^4$Department of Astronomy, University of Michigan, 500 Church St., Ann Arbor, MI 48109, USA\\
$^5$Institute for Astronomy, University of Edinburgh, Royal Observatory, Blackford Hill, Edinburgh EH9 3HJ, UK\\
$^6$Institute for Astronomy, University of Hawaii at Manoa, Honolulu, HI 96822, USA\\
}

\date{\today}

\begin{document} 
\maketitle 
\begin{abstract} 
We present the first spectroscopic analysis of the faint and compact stellar system Draco II (Dra~II, $M_V=-2.9\pm0.8$, $r_h=19^{+8}_{-6}\pc$), recently discovered in the Pan-STARRS1 $3\pi$ survey. The observations, conducted with DEIMOS on the Keck~II telescope, establish some of its basic characteristics: the velocity data reveal a narrow peak with 9 member stars at a systemic heliocentric velocity $\langle v_r\rangle=-347.6^{+1.7}_{-1.8}\kms$, thereby confirming Dra~II is a satellite of the Milky Way; we infer a velocity dispersion with $\sigma_{vr}=2.9\pm2.1\kms$ ($<8.4\kms$ at the 95\% confidence level), which implies $\log_{10}\left(M_{1/2}\right)=5.5^{+0.4}_{-0.6}$ and $\log_{10}\left((M/L)_{1/2}\right)=2.7^{+0.5}_{-0.8}$, in Solar units; furthermore, very weak Calcium triplet lines in the spectra of the high signal-to-noise member stars imply $\FeH<-2.1$, whilst variations in the line strengths of two stars with similar colours and magnitudes suggest a metallicity spread in Dra~II. These new data cannot clearly discriminate whether Draco II is a star cluster or amongst the faintest, most compact, and closest dwarf galaxies. However, the sum of the three --- individually inconclusive --- pieces of evidence presented here, seems to favour the dwarf galaxy interpretation.
\end{abstract}

\begin{keywords} Local Group -- galaxy: individual (Draco~II) --- galaxies: kinematics and dynamics
\end{keywords}

\section{Introduction}
Systematic surveys of the Milky Way surroundings with CCD photometry such as the Sloan Digital Sky Survey (SDSS), the Panoramic Survey Telescope and Rapid Response System~1 (Pan-STARRS1 or PS1), and the Dark Energy Survey (DES) have allowed for the discovery of numerous faint Milky Way satellites in the last decade \citep[e.g.,][]{willman05a,belokurov07a,bechtol15,laevens15b}. Despite sometimes reaching total luminosities of only $\sim10^3\lsun$ \citep{martin08b}, a significant fraction of these new discoveries are confirmed to be dynamically hotter than implied by their baryonic content alone and are thought to be the most dark-matter dominated dwarf galaxies to orbit the Milky Way \citep[e.g.,][]{martin07a,simon07,geha09,kirby13b}. Such systems are particularly valuable to both understand the faint-end of galaxy formation \citep[e.g.,][]{brown14} and hunt for dark matter annihilation signals as their properties are not expected to be strongly impacted by baryonic processes \citep[e.g.,][]{bonnivard15}.

However, without spectroscopic observations, assessing the nature of such stellar systems is rendered difficult by the apparent merging of the globular cluster and dwarf galaxy realms at the faint end. Although this effect is likely due in part to surface brightness limits in the current searches that translate to only faint and small stellar systems being bright enough to overcome detection limits \citep{koposov07,walsh09,drlica-wagner15}, it remains that disentangling currently observed globular clusters from dwarf galaxies can be challenging. Two such examples, Draco~II (Dra~II) and Sagittarius~II, are shown in our recent presentation of three faint systems found in the PS1 $3\pi$ survey \citep{laevens15b}. With a total magnitude of $M_V=-2.9\pm0.8$ (or $\lsun=10^{3.1\pm0.3}$) and a half-light radius of only $r_h=19^{+8}_{-6}\pc$, Dra~II is a system whose properties are similar to those of the dwarf galaxy Segue~1 \citep{geha09,simon11}, but whose size is smaller than any confirmed dwarf galaxy.

In this letter, we analyze the first spectroscopic observations of Dra~II with the DEIMOS multi-object spectrograph on Keck~II \citep{faber03}. The measured velocities confirm that Dra~II is a Milky Way satellite and show a marginally resolved velocity dispersion. The metallicity properties of the system further hint that Dra~II is likely a dwarf galaxy. We present our observations and data in Section~2, perform the analysis of the data set in Section~3, and conclude in Section~4.

\section{Observations and data}
One DEIMOS mask was observed on the night of July 17, 2015, placed close to the center of Dra~II in such a way to optimise the number of high-priority bright candidate members. The priorities were set as both a function of spatial location (higher priority towards the center of the system) and location in the colour-magnitude diagram (CMD). All targets are selected using the PS1 photometry and higher priorities are given to potential main-sequence (MS), main-sequence turn off (MSTO) and red-giant-branch (RGB) stars selected to follow an isochrone that best reproduces the CMD features of Dra~II. The mask was drilled with $0.7''$ slits.

Observations were taken following our usual routine \citep[e.g.,][]{martin14b} for a total of 3,600s, split into three 1,200s sub-exposures, under good conditions (50\% humidity, $0.7''$ seeing). We further observed NeArKrXe calibrations through the slit mask after the science frames at the same location on sky. The chosen grating has 1200~lines/mm and covers the wavelength range 6600--9400\AA, with a spectral resolution of $\sim0.33$\AA\ per pixel.

We process the raw spectra through our own pipeline that we developed over the years to specifically handle DEIMOS data. The details of the pipeline, which focuses on the Calcium triplet region, are given by \citet{ibata11a}, to which we add another calibration step using the Fraunhofer A band in the range 7595--7630\AA\ in order to perform small telluric corrections \citep{martin14b}. The signal to noise per pixel (S/N) of the reduced spectra in the Ca triplet region is typical 30/8 at $i_\mathrm{P1}=18.0/20.0$.

For a cold stellar system like Dra~II, it is particularly important to assess the level of systematics on the measured velocity uncertainties. DEIMOS is known to yield a small level of systematics that cannot be entirely explained from properly tracking the sources of noise in the spectra. These systematics are likely due to minute misalignments of stars in the slits and can only be constrained through repeat measurements of observations and/or a comparison with reference radial velocities. \citet{ibata11a} conducted such a comparison for high signal-to-noise DEIMOS spectra of the NGC~2419 globular cluster, observed under very similar conditions ($0.7''$ slits and $\sim0.7''$ seeing) and processed through our pipeline. The comparison was made with more accurate HIRES observations of 7 stars and yielded an uncertainty floor of $2.25\kms$, which we add in quadrature to the velocity uncertainties measured from the spectra.%\footnote{We favor the uncertainty floor of \citet{ibata11a} over the one we measured in \citet[$3.4\pm0.5\kms$]{martin14b} as this latter measurement was performed on a sample of RGB stars near the tip of the RGB. Such stars are known to suffer from `jitter' in their atmosphere, which produces velocity differences of up to a few $\kms$ for repeat measurements \citep[e.g.,][]{carney08}. It is therefore very likely that the \citet{martin14b} uncertainty floor is inflated by this effect and not adequate for the MSTO and faint RGB stars we observe in Dra~II. It should also be noted that the \citet{ibata11a} measurement is in perfect agreement with that of \citet{simon07}, despite stemming from different data sets and reduction pipelines.}.

After culling stars with low signal-to-noise ($S/N<3$ per pixel) and velocity uncertainties higher than $15\kms$, we converge on a final sample of 34 stars with good radial velocity measurements. The properties of the 9 Dra~II member stars (see below) are listed in Table~1 and the properties of field stars are available online.

All velocities given in this letter are heliocentric radial velocities, except when indicated otherwise.
 
 \begin{table*}
\caption{\label{data}Properties of observed stars.}
\begin{tabular}{lllccccccccc}
\# & RA & Dec & $g_\mathrm{P1}$ & $\delta_{g_\mathrm{P1}}$ & $i_\mathrm{P1}$ & $\delta_{i_\mathrm{P1}}$ & Member & $v_r$ & $\delta_{vr}$ & S/N & Tentative $\FeH$\\
 & (ICRS) & (ICRS) &  &  &  &  &  & ($\kms$) & ($\kms$) &  (per pixel) & \\
\hline
%1 &  238.3453674 &    64.5739746 & 19.139 & 0.009  & 18.684 & 0.007 & N & $-123.7$ & 2.4 & 25.4 \\
2 &  238.2920837 &    64.5601120 & 19.397 & 0.012  & 18.865 & 0.006 & Y & $-344.1$ &    2.4 & 25.2  & $-2.3\pm0.1$\\
%3 &  238.2310028 &    64.5967789 & 17.811 & 0.006  & 17.190 & 0.004 & N & $  -3.9$ &    2.3 & 53.8 \\
4 &  238.2274933 &    64.5717239 & 19.888 & 0.015  & 19.419 & 0.010 & Y & $-349.8$ &    3.0 & 14.6 \\
5 &  238.2233734 &    64.5534134 & 20.101 & 0.019  & 19.771 & 0.013 & Y & $-354.4$ &    3.3 & 11.3 \\
%6 &  238.2153320 &    64.6064758 & 15.487 & 0.003  & 15.071 & 0.003 & N & $ -55.2$ &    2.2 & 69.4 \\
%7 &  238.2096710 &    64.6198349 & 20.352 & 0.031  & 19.454 & 0.011 & N & $-183.6$ &    4.9 &  8.1 \\
%8 &  238.1847992 &    64.6154175 & 17.977 & 0.006  & 17.179 & 0.003 & N & $ -36.4$ &    2.4 & 43.8 \\
9 &  238.1757050 &    64.5701370 & 19.994 & 0.021  & 19.391 & 0.010 & Y & $-343.1$ &    3.0 & 12.3 \\
10 &  238.1514130 &    64.6053925 & 19.528 & 0.014  & 18.975 & 0.007 & Y & $-346.7$ &    3.0 & 15.7 &  $-3.5^{+0.5}_{-0.8}$\\
%11 &  238.1316223 &    64.5645523 & 18.882 & 0.008  & 18.274 & 0.005 & N & $-135.6$ &    2.3 & 27.3 \\
%12 &  238.1148682 &    64.6088638 & 17.916 & 0.006  & 16.969 & 0.003 & N & $-122.3$ &    2.3 & 58.6 \\
%13 &  238.1112518 &    64.5757751 & 17.459 & 0.005  & 16.705 & 0.003 & N & $ -68.7$ &    2.3 & 52.5 \\
%14 &  238.0767517 &    64.5960846 & 20.510 & 0.032 & 19.190 & 0.017 & N & $ -95.5$ &    3.2 & 11.5 \\
%15 &  238.0709229 &    64.5575027 & 17.602 & 0.005 & 16.856 & 0.003 & N & $-281.5$ &    2.3 & 55.6 \\
%16 & 238.5662842 &    64.5052795 & 19.038 & 0.009 & 18.468 & 0.005 & N & $-227.4$ &    2.4 & 22.4 \\
%17 &  238.5580902 &    64.5661392 & 21.795 & 0.079  & 20.145 & 0.015 & N & $ -44.2$ &    3.4 & 10.6 \\
%18 &  238.5460358 &    64.5298080 & 20.464 & 0.031  & 19.986 & 0.015 & N & $-159.3$ &    4.1 &  8.1 \\
%19 &  238.5143738 &    64.5734711 & 17.957 & 0.007  & 17.481 & 0.004 & N & $ -81.2$ &    2.3 & 39.5 \\
%20 &  238.5104980 &    64.5553360 & 19.924 & 0.017  & 18.300 & 0.005 & N & $ -56.4$ &    2.4 & 32.5 \\
%21 &  238.5098267 &    64.5115814 & 15.939 & 0.005  & 15.297 & 0.003 & N & $ -24.8$ &    2.2 & 63.3 \\
%22 &  238.4972992 &    64.5178299 & 18.134 & 0.006  & 16.318 & 0.002 & N & $ -11.8$ &    2.2 & 57.2 \\
%23 &  238.4397430 &    64.5676422 & 18.177 & 0.008  & 16.947 & 0.003 & N & $ -65.2$ &    2.3 & 50.8 \\
%24 &  238.4327087 &    64.5393295 & 20.737 & 0.035  & 19.264 & 0.009 & N & $-103.0$ &    2.5 & 19.6 \\
25 &  238.4129944 &    64.5798874 & 22.436 & 0.149  & 21.583 & 0.081 & Y & $-354.3$ &    5.7 &  3.4 \\
%26 &  238.3889160 &    64.5784988 & 20.544 & 0.031  & 19.453 & 0.009 & N & $ -99.6$ &    2.5 & 16.5 \\
27 &  238.2976685 &    64.5859756 & 21.108 & 0.056  & 20.632 & 0.025 & Y & $-354.8$ &    7.5 &  5.3 \\
%28 &  238.2884216 &    64.5899429 & 21.424 & 0.078  & 19.774 & 0.013 & N & $ -65.2$ &    2.7 & 13.4 \\
%29 &  238.2546234 &    64.5418854 & 20.605 & 0.026  & 20.130 & 0.017 & N & $  10.1$ &    3.3 &  8.4 \\
30 &  238.2506714 &    64.5479965 & 21.535 & 0.063  & 21.295 & 0.044 & Y & $-344.7$ &    7.0 &  3.2 \\
%31 &  238.2260895 &    64.5978851 & 19.629 & 0.016  & 18.627 & 0.007 & N & $ -86.0$ &    2.4 & 27.0 \\
32 &  238.2179565 &    64.5957489 & 20.789 & 0.050  & 20.569 & 0.024 & Y & $-343.0$ &    7.3 &  3.8 \\
%33 &  238.2109985 &    64.5784760 & 20.096 & 0.024  & 18.626 & 0.006 & N & $  -5.4$ &    2.4 & 25.8 \\
%34 &  238.1220856 &    64.6019974 & 18.954 & 0.008  & 17.005 & 0.003 & N & $  -1.9$ &    2.3 & 53.1 \\
\end{tabular}
\emph{Note:} The full table, including non-member foreground Milky Way stars, is available online.
\end{table*}

\section{Results}
\subsection{Velocities}
\begin{figure}
\begin{center}
\includegraphics[width=0.75\hsize,angle=270]{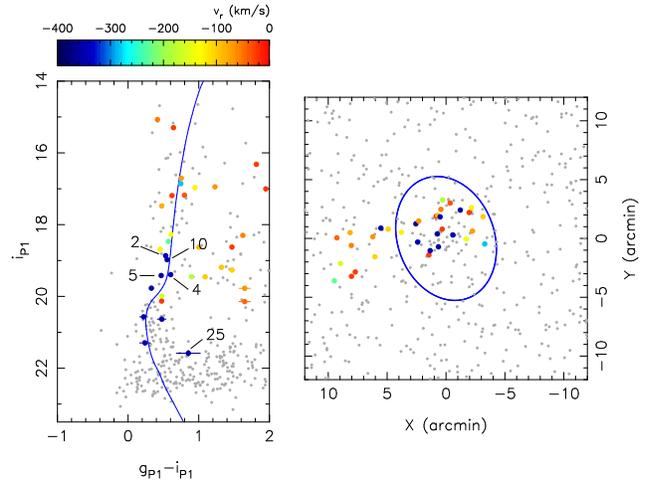}
\caption{\label{CMD}\emph{Left:} PS1 CMD of stars within $3 r_h$ of the center of Dra~II. Stars in our spectroscopic sample are colour-coded by their heliocentric velocities whilst stars without spectroscopy are shown in gray. The error bars represent the photometric uncertainties for the stars with velocities. The 9 Dra~II member stars appear as dark blue. The blue line is an isochrone with the properties assigned to the stellar system by \citet[$13\Gyr$, $\FeH=-2.2$, $m-M = 16.9$]{laevens15b}. The 4 member stars whose spectra are displayed in Figure~\ref{spectra} are labeled. \emph{Right:} Distribution of PS1 stars in the region of Dra~II. The colour-coding is the same as in the left-hand panel. The blue ellipse delineates the region within $2r_h$ of the system's center as determined by \citet{laevens15b}.}
\end{center}
\end{figure}

\begin{figure}
\begin{center}
\includegraphics[width=0.75\hsize,angle=270]{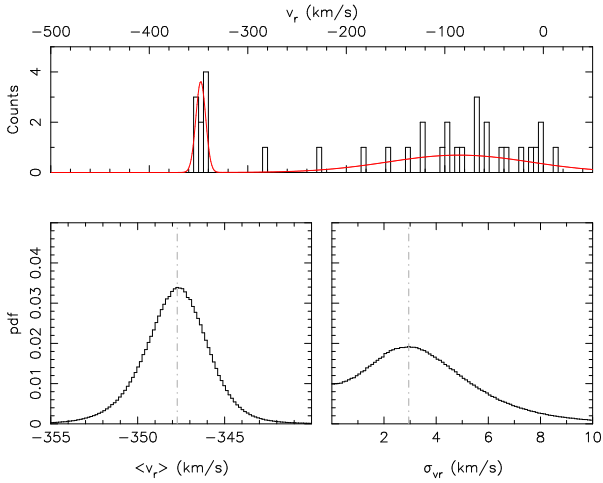}
\caption{\label{velFit}\emph{Top:} Heliocentric velocity distribution of the spectroscopic sample. The cold velocity peak at $v_r\sim-350\kms$ is produced by Dra~II stars. The red line displays the best fit to the velocity distribution, convolved by the median velocity uncertainty. \emph{Bottom:} Probability distribution functions of the two fit parameters relevant to Dra~II: the systemic velocity of the satellite (left) and its velocity dispersion (right). The gray dashed lines indicate the mode of the distributions.}
\end{center}
\end{figure}

The location of the 34 sample stars in the Dra~II CMD and on the sky is displayed in Figure~\ref{CMD}, colour-coded by their heliocentric velocities. Already, one can note a sub-sample of stars that track the CMD features of Dra~II at large negative velocities. This is confirmed by the velocity distribution of the whole sample, presented in the top panel of Figure~\ref{velFit}, that clearly exhibits a cold velocity peak near $v_r\sim-350\kms$, within the expected range for a Milky Way satellite. The 9 stars that compose the velocity peak are those overlaid in dark blue in Figure~\ref{CMD}. All 9 stars are quite faint and belong to the stellar system's MS, MSTO, or low RGB. The isochrone of an old ($13\Gyr$) and metal-poor ($\FeH=-2.2$) stellar population at the distance of Dra~II \citep[$m-M\sim16.9$;][]{laevens15b} is shown for comparison.

Despite our selection of a large number of potential (brighter) RGB stars, we did not uncover a single star with $i_\mathrm{P1}<18.5$ in Dra~II. On the other hand, the member stars are, as expected, located towards the center of the system. All but one member star lie in the ellipse delimiting the region within $2r_h$.

We fit the velocity distribution by a model composed of the sum of 2 Gaussian functions representing the Dra~II signal and the MW foreground contamination. Following the probabilistic framework presented in \citet{martin14b}, which takes the individual velocity uncertainties into account, yields the Dra~II systemic velocity, $\langle v_r\rangle=-347.6^{+1.7}_{-1.8}\kms$ or $\langle v_{r,\mathrm{gsr}}\rangle \simeq -180\kms$, and its velocity dispersion, $\sigma_{vr}=2.9\pm2.1\kms$, with a 95\% confidence limit of $8.4\kms$. The probability distribution functions are also shown in Figure~\ref{velFit} for these two parameters%\footnote{Performing the same analysis on the sample of 6 member stars independently reduced through the \citet{kirby15} pipeline yields an unresolved velocity dispersion that is nevertheless consistent with $\sigma_{vr}<4.8\kms$ at the 90\% confidence level.}
. The set of parameters that maximises the likelihood function is used to build the velocity model, which can be seen in the top panel of the figure after convolution by the median uncertainty. It compares very favourably with the velocity distribution.

One may question the membership of the faintest star in the sample, star 25, as it is rather red compared to the isochrone shown in the left-hand panel of Figure~\ref{CMD}. It happens to also be the outermost star with a high membership probability, appearing as the only dark blue data point beyond the $2r_h$ ellipse in the right-hand panel of Figure~\ref{CMD}. However, the faint magnitude of this star translates into a large velocity uncertainty ($5.7\kms$) and, consequently, keeping or removing it from the sample of members does not impact the inference on the velocity properties of Dra~II.

\begin{figure}
\begin{center}
\includegraphics[width=0.65\hsize,angle=270]{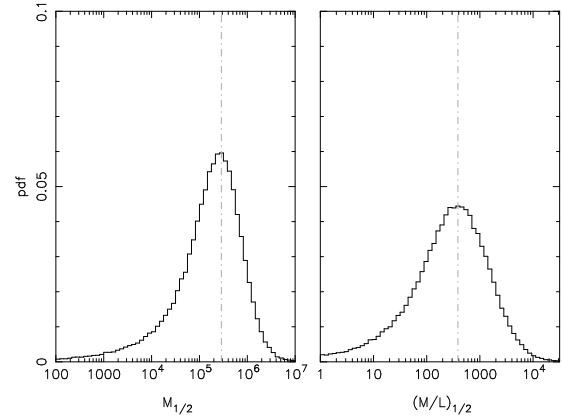}
\caption{\label{M_ML} Probability distribution functions of the dynamical mass of Dra~II within its 3-dimensional half-light radius ($M_{1/2}$; left) and of its mass-to-light ratio within the same radius ($(M/L)_{1/2}$). The gray dashed lines indicate the mode of the distributions.}
\end{center}
\end{figure}

Although our contraints on the velocity dispersion are weak, it is marginally resolved and close to values measured for faint MW dwarf galaxies such as Segue~1 \citep[$3.9\pm0.8\kms$]{simon11}. Using equation (1) of \citet{wolf10}, we can estimate the mass within the 3-dimensional half-light radius, $M_{1/2}$, via the half-light radius $r_h$: $M_{1/2}\simeq930r_h\sigma_{vr}^2\msun$. Randomly drawing values from the pdfs of $\sigma_{vr}$ from above and $r_h$ from \citet{laevens15b} yields $\log_{10}\left(M_{1/2}\right)=5.5^{+0.4}_{-0.6}$ and $\log_{10}\left((M/L)_{1/2}\right)=2.7^{+0.5}_{-0.8}$, in Solar units (Figure~\ref{M_ML}).

If Dra~II were a stellar system in equilibrium and binaries had no impact, one would expect a velocity dispersion of order $\sim0.3\kms$. Therefore, even though we cannot rule out that the large dynamical mass we measure could be a statistical fluctuation, we find marginal evidence that Dra~II is hotter than would be implied solely by its baryonic content, hinting that it could be a dark-matter dominated dwarf galaxy. However, more velocities are required to strengthen the velocity dispersion measurement and, in particular, assess the impact of binary stars \citep{mcconnachie10,simon11}.

Unrelated to Dra~II member stars, we note in passing that the distribution of MW foreground contaminants tends to favour negative velocities. In particular, 2 stars have velocities below $-220\kms$, which is quite unexpected (a $3\sigma$ deviation from expectations) and could point towards the presence of halo stellar substructure along this line of sight.

\subsection{Metallicities}
\begin{figure}
\begin{center}
\includegraphics[width=\hsize]{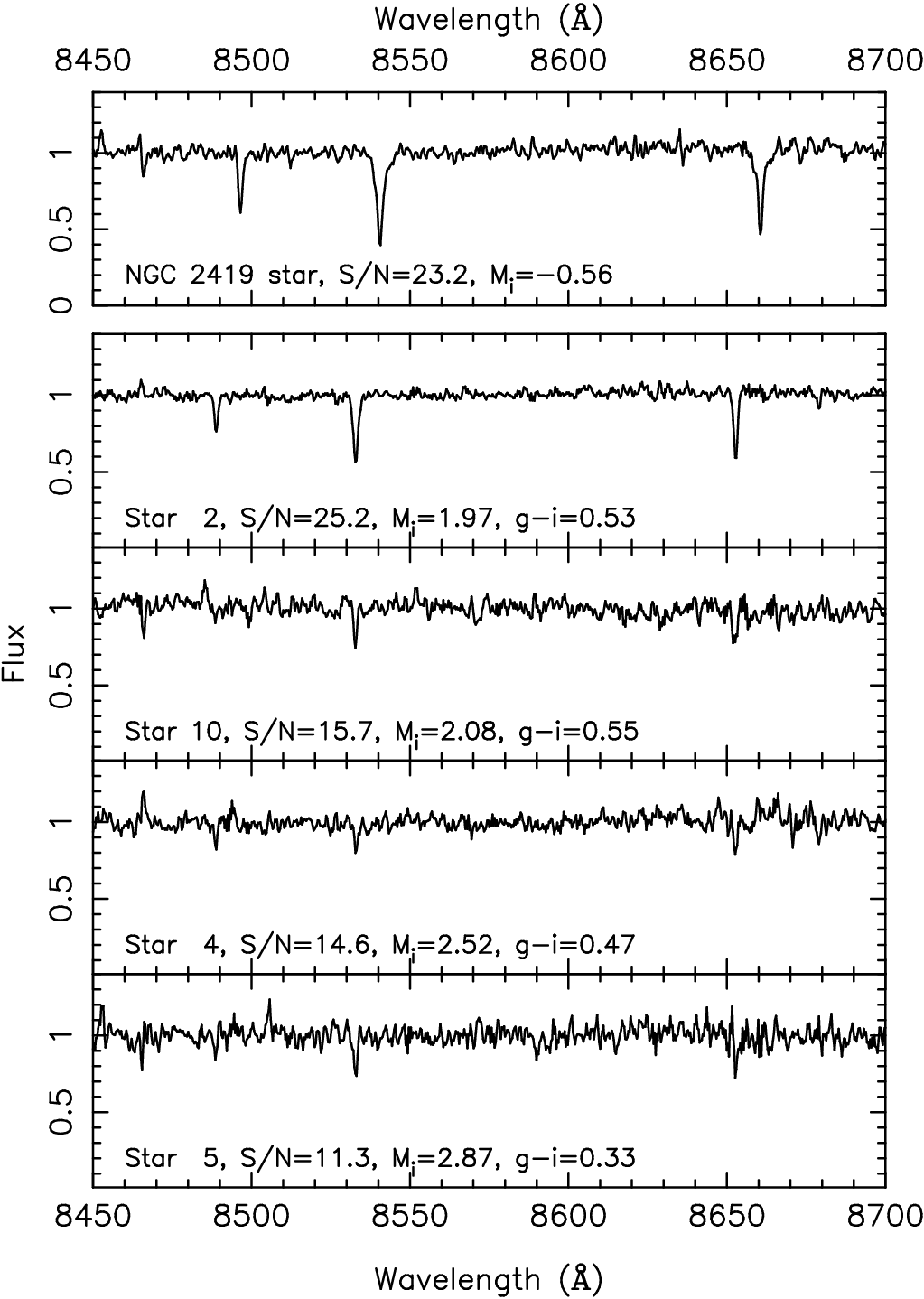}
\caption{\label{spectra}Comparison of the spectrum of a faint RGB star from NGC~2419 (top), taken from the \citet{ibata11a} sample, with the spectra of the 4 Dra~II member stars with the highest $S/N$ (bottom four panels). The spectra are smoothed with a 3-pixel boxcar kernel. The wavelength coverage shown includes the Ca triplet lines, clearly visible in the spectrum of the NGC~2419 star and weaker in star~2 of the Dra~II. These three strong lines are almost buried in the noise of the other 3 spectra. Stars~2 and 10 display different line depths, despite having similar colours and magnitudes, implying a metallicity dispersion in the system.}
\end{center}
\end{figure}

The absence of bright RGB stars amongst the 9 Dra~II member stars limits our ability to accurately measure the metallicity of the system. However, we note that the high S/N member stars exhibit particularly weak Ca triplet lines, which implies a very low $\FeH$ metallicity. In Figure~\ref{spectra}, we compare the spectra of Dra~II member stars 2, 10, 4, and 5 (the 4 stars with the highest $S/N$) with the spectrum of a star that belongs to NGC~2419 ($\FeH=-2.1$; \citealt{cohen10}), observed by \citet{ibata11a} with the same instrumental set up. The brightest 4 Dra~II member stars all show weaker lines than the metal-poor NGC~2419. In fact, the Ca triplet lines of stars~10, 4, and 5 are barely distinguishable from the noise in the spectra, despite $10<S/N<15$.

At this stage, a word of caution is necessary as the NGC~2419 star is significantly brighter ($M_i = -0.6$) than the Dra~II stars ($M_i \simeq +2.0$). As such, it is expected that the CaT lines of the Dra~II stars should be less pronounced for the same overall metallicity. In addition, the \citet{starkenburg10} relation between the equivalent widths of theses lines and the metallicity of their stars has not been calibrated fainter than the horizontal branch ($M_i\simeq0.9$) so we are loath to blindly use this relation to quote $\FeH$ values for these stars. However, \citet{leaman13} have demonstrated in the case of the metal-poor globular cluster NGC~7078 that the \citet{starkenburg10} relation is consistent with observations down to at least $\sim2$ magnitudes below the horizontal branch. Figure~1 of the Leaman et al. paper shows that the CaT equivalent width\footnote{In the following, CaT equivalent widths and their uncertainties are estimated by fitting Gaussian functions to the second and third Ca lines. These are then summed into a global equivalent width, as per \citet{starkenburg10}.} difference between the metal-poor NGC~2419 star ($2.94\pm0.19$\AA) and star 2 from Dra~II ($1.63\pm0.11$\AA) is driven mainly by the change of $\log g$ along the RGB and that this star is consequently as metal-poor as NGC~2419. It further implies that star~10, with a CaT equivalent width of only $0.75\pm0.21$ is significantly more metal-poor than NGC~2419.

Furthermore, the stark difference between the spectra of stars~2 and 10, which must have very similar stellar parameters as they are confirmed Dra~II member stars with almost identical colours and magnitudes ((0.53,18.87) and (0.55,18.98)), implies that these two member stars have significantly different metallicities (a $4.5\sigma$ difference in the equivalent width measurements). Therefore, we are left to conclude that Dra~II is not only a metal-poor system, but also has a metallicity dispersion. Going further and estimating tentative metallicity values for stars 2 and 10 via equation (A.1) of \citet{starkenburg10} for I-band magnitudes \footnote{These I-band magnitudes are inferred from $i_\mathrm{P1}$ via the \citet{tonry12} colour equations} yields $\FeH=-2.3\pm0.1$ and $-3.5^{+0.5}_{-0.8}$, respectively, in good agreement with the conclusions of the comparison with the NGC~2419 star.

Taking the dwarf galaxy luminosity--metallicity relation of \citet{kirby13} at face value, one would expect $\FeH\sim-2.6$ for a system of Dra~II's overall luminosity, which is compatible with our findings. In addition, only dwarf galaxies exhibit metallicity dispersions at these magnitudes \citep{willman12}. Therefore, the low metallicity and the metallicity dispersion implied by our analysis lead us to conclude that, independently of the kinematics, Dra~II is likely a dwarf galaxy and not a globular cluster.

\section{Conclusions}

In this letter, we performed a first spectroscopic study of the Dra~II stellar system recently discovered in the PS1 $3\pi$ survey, which establishes some of its basic properties:

\begin{enumerate}
\item A systemic velocity of $\langle v_{r,\mathrm{gsr}}\rangle\simeq-180\kms$ confirms it is indeed a satellite of the Milky Way.
\item The inferred velocity dispersion of Dra~II is $\sigma_{vr} = 2.9\pm2.1\kms$. Combined with the size of the system, it implies a mass-to-light ratio within the 3-dimensional half-light radius, $\log\left((M/L)_{1/2}\right) = 2.7^{+0.5}_{-0.8}$, that is hard to reconcile with a baryonic system in equilibirum.
\item The Calcium triplet lines of Dra~II member stars imply that the system is more metal-poor than the NGC~2419 globular cluster, i.e. $\FeH<-2.1$. The quasi-absence of CaT lines in some of the member stars implies that the systematic metallicity of the system could be significantly more metal-poor than this, as expected from the dwarf galaxies' luminosity--metallicity relation.
\item Two Dra~II member stars with similar colours and magnitudes have significantly different equivalent widths (a $4.5\sigma$ difference), which implies a metallicity dispersion in Dra~II. No $\FeH$ dispersion has ever been observed in low-luminosity globular clusters, but is commonly observed in dwarf galaxies.
\end{enumerate}

None of the above measurements or arguments by itself can discriminate whether Dra~II is a star cluster or a dwarf galaxy. Taken together, however, these measurements favour on balance the interpretation that this system is amongst the faintest, most compact, and closest dwarf galaxies ($r_h=19^{+8}_{-6}\pc$, $L_V=10^{3.1\pm0.3}\lsun$, and $D_\mathrm{Helio}\sim20\kpc$) and a target of choice for both the study of the faint end of galaxy formation and for searches of indirect dark-matter detections.

\section*{Acknowledgments}
We wish to thank Else Starkenburg for discussions regarding her calibration of the RGB equivalent-width--metallicity relation. B.P.M.L. acknowledges funding through a 2012 Strasbourg IDEX (Initiative d'Excellence) grant, awarded by the University of Strasbourg. N.F.M. and B.P.M.L. gratefully acknowledge the CNRS for support through PICS project PICS06183. H.-W.R. acknowledges support by the DFG through the SFB 881 (A3).

The data presented herein were obtained at the W.M. Keck Observatory, which is operated as a scientific partnership amongst the California Institute of Technology, the University of California and the National Aeronautics and Space Administration. The Observatory was made possible by the generous financial support of the W.M. Keck Foundation. The authors wish to recognise and acknowledge the very significant cultural role and reverence that the summit of Mauna Kea has always had within the indigenous Hawaiian community.  We are most fortunate to have the opportunity to conduct observations from this mountain.

The Pan-STARRS1 Surveys (PS1) have been made possible through contributions by the Institute for Astronomy, the University of Hawaii, the Pan-STARRS Project Office, the Max-Planck Society and its participating institutes, the Max Planck Institute for Astronomy, Heidelberg and the Max Planck Institute for Extraterrestrial Physics, Garching, The Johns Hopkins University, Durham University, the University of Edinburgh, the Queen's University Belfast, the Harvard-Smithsonian Center for Astrophysics, the Las Cumbres Observatory Global Telescope Network Incorporated, the National Central University of Taiwan, the Space Telescope Science Institute, and the National Aeronautics and Space Administration under Grant No. NNX08AR22G issued through the Planetary Science Division of the NASA Science Mission Directorate, the National Science Foundation Grant No. AST-1238877, the University of Maryland, Eotvos Lorand University (ELTE), and the Los Alamos National Laboratory.

\newcommand{\mnras}{MNRAS}
\newcommand{\pasa}{PASA}
\newcommand{\nat}{Nature}
\newcommand{\araa}{ARAA}
\newcommand{\aj}{AJ}
\newcommand{\apj}{ApJ}
\newcommand{\apjl}{ApJ}
\newcommand{\apjs}{ApJSupp}
\newcommand{\aap}{A\&A}
\newcommand{\aaps}{A\&ASupp}
\newcommand{\pasp}{PASP}

% Bibtex will create a .bbs file in the directory and before sending to the editor, I should replace the bibliography call by this file.

\end{document}